\documentclass[journal,9pt]{IEEEtran}
\usepackage{cite}
\ifCLASSINFOpdf

\usepackage[pdftex]{graphicx}
\else
\fi
\usepackage[cmex10]{amsmath}
\usepackage{url}
\usepackage{newcent}
\begin{document}
%\title{Impact of Advanced Signal Processing (Wavelength and Modulation Conversion) on Elastic Optical Networks Using MILP}
\title{Impact of Wavelength and Modulation Conversion on Transluscent Elastic Optical Networks Using MILP}
%
%
% author names

\author{X. Wang, M.~Brandt-Pearce and~S.~Subramaniam% <-this % stops a space
%\thanks{Manuscript received April 19, 2005; revised January 11, 2007.}
\thanks{X. Wang and M. Brandt-Pearce are with the Charles L. Brown Department of Electrical and Computer Engineering,
University of Virginia, Charlottesville, Virginia 22904 USA (e-mail: mb-p@virginia.edu). }%
\thanks{S. Subramaniam is with the Department
of Electrical and Computer Engineering, The George Washington University, Washington,
DC, 20052 USA (e-mail: mb-p@virginia.edu).}}% <-this % stops a space

\maketitle

\begin{abstract}
%\boldmath
Compared to legacy wavelength division multiplexing networks, elastic optical networks (EON) have added flexibility to network deployment and management. EONs can include previously available technology, such as signal regeneration and wavelength conversion, as well as new features such as finer-granularity spectrum assignment and modulation conversion. Yet each added feature adds to the cost of the network.  In order to quantify the potential benefit of each technology, we present a link-based mixed-integer linear programming (MILP) formulation to solve the optimal resource allocation problem. We then propose a recursive model in order to either augment existing network deployments or speed up the resource allocation computation time for larger networks with higher traffic demand requirements than can be solved using an MILP.  We show through simulation that systems equipped with  signal regenerators or wavelength converters require a notably smaller total bandwidth, depending on the topology of the network. We also show that the suboptimal recursive solution speeds up the calculation and makes the running-time more predictable, compared to the optimal MILP.
\end{abstract}

\begin{IEEEkeywords}
Elastic optical networks, spectral resource allocation, mixed integer linear programming, regeneration placement, modulation conversion.
\end{IEEEkeywords}

% \fi
\IEEEpeerreviewmaketitle

%\section{Introduction}
% by the rest of the first.
% normal text:
% and "his" in lower case to complete the first word.
%\IEEEPARstart{T}{}his demo file is intended to serve as a
%``starter file'' for Journal of Optical Communications and
%Networking papers produced under \LaTeX\ using IEEEtran.cls
%version 1.7 and later.
%% You must have at least 2 lines in the paragraph with the drop letter
%% (should never be an issue)
%I wish you the best of success.

\section{Introduction}
\IEEEPARstart{I}{}ncreasing traffic volume and growing heterogeneity of bandwidth requirements have pushed the development of optical transport networks. Using wavelength division multiplexing (WDM) technology, spectrum usage has greatly increased by allowing multiple-line-rates and traffic grooming \cite{Wang-2012}. Yet WDM is unable to handle increasing traffic heterogeneity because of the coarse wavelength grid employed. Elastic optical networks (EONs), on the other hand, provide flexibility in both bandwidth assignment (using sub-channel granularity and super-channel assignments) and lightpath reconfigurability not available in WDM. As the technology matures, additional functionality such as modulation selection and conversion can be added, with the hope of further increasing the spectral efficiency. When major additions in physical layer features are being considered, the network design should be re-examined to determine the realized benefit gained by their implementation. This paper presents an optimal routing, regeneration, and spectrum allocation formulation that is then used to evaluate the merit of wavelength and modulation conversion on EONs affected by physical layer impairments.

The design of transport networks includes the placement and assignment of all physical resources, such as optical fiber and electronic devices (transponders, high speed optical-electrical-optical conversion circuits, etc.).  The goal is usually to minimize the capital expenditure while fulfilling certain traffic accommodation expectations. One common way to solve this problem is to address it as a multi-commodity assignment by pairing the physical resources with traffic demands in order to minimize the resources used by each demand. For example, a traffic demand can be assigned the shortest route in order to reduce the cost. Such design principles have been used to develop many heuristic algorithms for network design \cite{Erlebach-2003, Garcia-2007, Zyane-2014}. Although these algorithms are computationally simple, they often yield poor performance when the problem becomes complex and consideration cannot be given to all influencing factors. Another approach is to formulate the resource allocation as an optimization problem with physical and network layer constraints and use linear programming (LP) to solve it. The available network resources become the LP design variables. Unlike arbitrary multi-commodity assignment problems, network design often requires its variables to be integer or Boolean, which leads to a mixed-integer linear programming (MILP) formulation. This significantly increases the computational complexity, not providing an approach that can scale to address larger networks.  However, for small networks and few traffic demands, the MILP can be solved in reasonable time, and results in an optimal solution, unlike heuristic algorithms. It does this without requiring a complete understanding of the relationship between the multiple design factors, as heuristic algorithms often do \cite{Talebi-2014}.

In this paper we develop an MILP design method for EONs.  Our formulation can implement modulation scheme selection, mid-lightpath modulation conversion (MC) and/or wavelength conversion (WC),  and regeneration circuit allocation (to satisfy either a quality of service constraint or conversion function). MILP has previously been used to solve the resource assignment optimization problem in optical networks \cite{Klinkowski-2011, Klinkowski-2012, Albert-2014}. However, to the best of our knowledge, no published MILP solution has included these flexibilities in an optimal way for designing EONs.

Acknowledging the limitations of the MILP approach for solving realistically-scaled problems due to its computational complexity, we envision the following two direct uses for our model.  The MILP can be solved for a small network to quantify the potential benefit that can be obtained by implementing a particular feature, such as modulation conversion, without introducing artificial limitations imposed by a suboptimal resource allocation algorithm. Our approach can also be used on realistic-size networks to solve for the optimal resource allocation of only a few traffic demands at a time. For the off-line resource allocation problem (static network), we can partition the whole traffic matrix into small sub-matrices, and solve the assignment problem for the sub-matrices in a sequential manner. By doing this, we are able to greatly reduce the overall execution time, in exchange for obtaining a suboptimal solution. In the paper we discuss the tradeoff between complexity and optimality for this approach that we call the recursive solution. For dynamic networks, we can use the recursive MILP to allocate resources for one or a few new connection requests given the current state of the deployed network, as we proposed for WDM systems in \cite{Wang-2012b}.

The rest of the paper is organized as follows: Section \ref{sec:network} introduces the network and node structure of our model and describes the advanced signal processing functionalities for EON that we consider; Section \ref{sec:ilp} explains how we implement the new functionalities with our MILP formulation; Section \ref{sec:recursive} develops our recursive MILP implementation that balances optimality and complexity; Section \ref{sec:simulation} presents numerical simulation results collected by solving the design problem using our formulation. Finally, conclusions are given in Section \ref{sec:conclusion}.

%------------------------------------------Network Description ---------------------------------------%
% 1. Network characteristics
% 2. Node structure
% 3. Traffic characteristics
% This section is additional introduction of the basis of our assumptions like multiple modulation schemes, signal regeneration and wavelength conversion, modulation conversion. This section requires many references.
\section{Network Description} \label{sec:network}
\begin{figure}[!tb]%[htdp]
   \centering
        \includegraphics[width=3in]{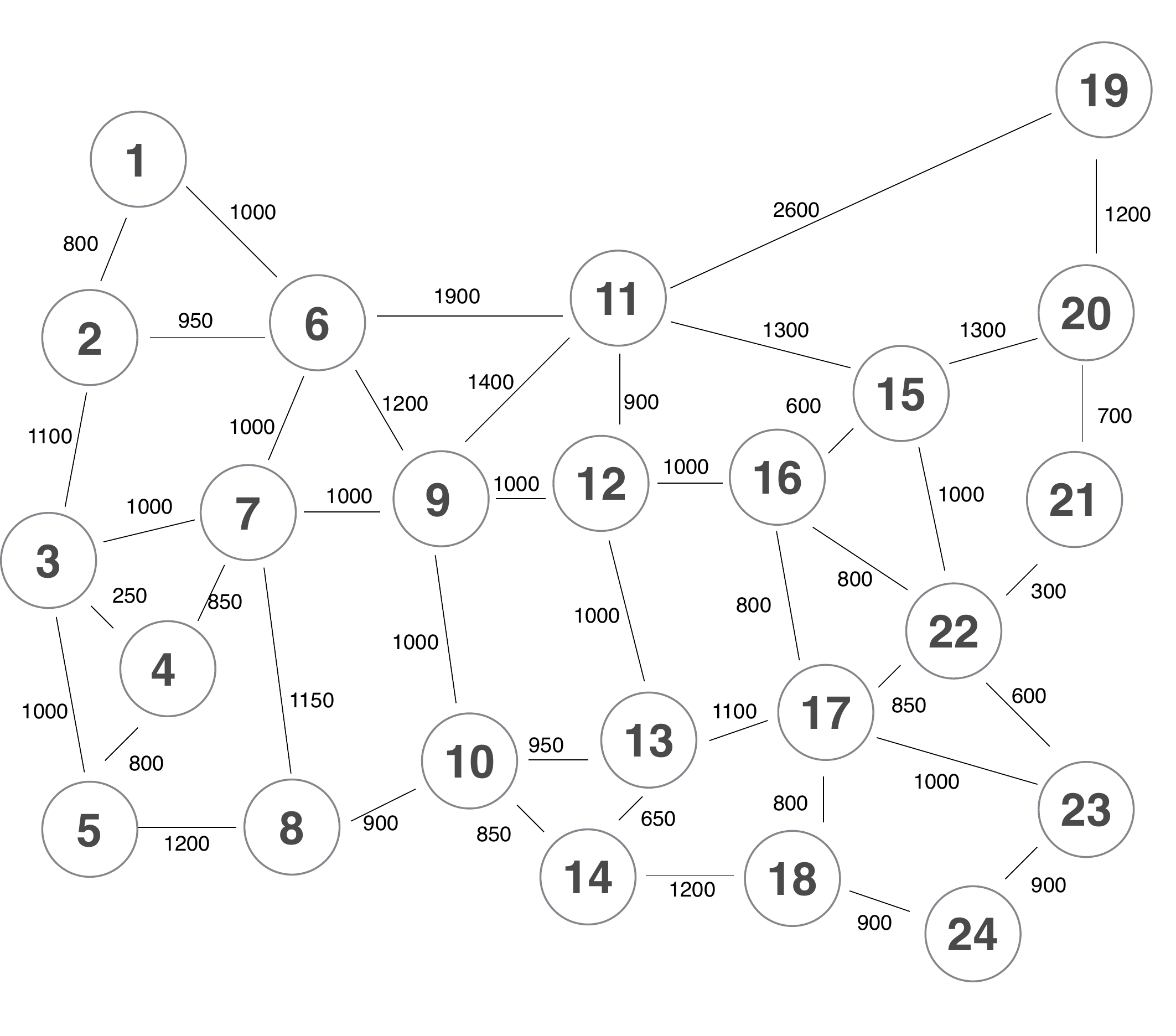}
    \caption{NSF-24 network. The number on each link represents the physical length of the link in km.}
    \label{fig:topologyNSF}
\end{figure}

\begin{figure}[!tb]%[htdp]
   \centering
        \includegraphics[width=3in]{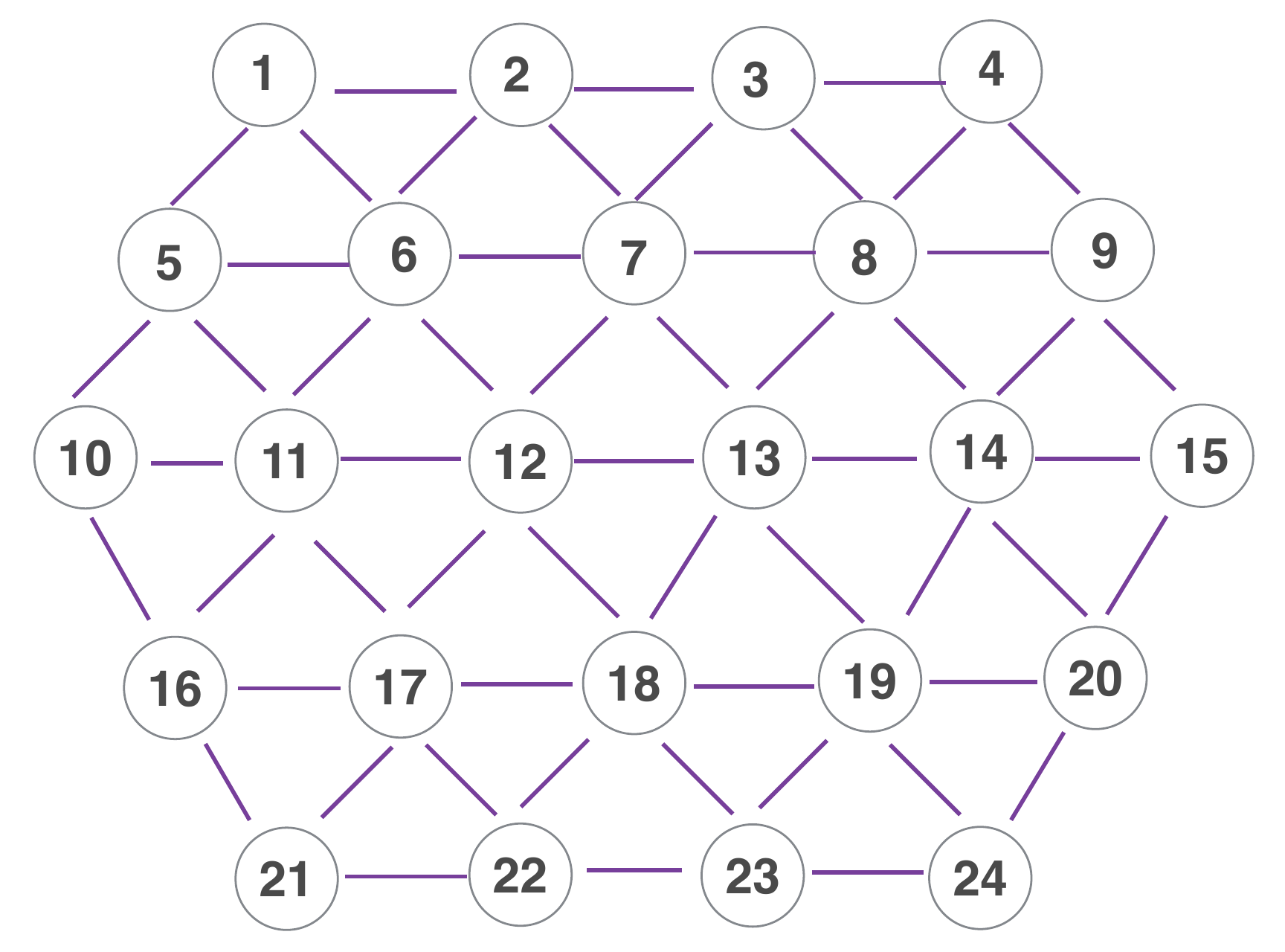}
    \caption{Symmetric-24 network}
    \label{fig:topologySymmetric}
\end{figure}

We are interested in long-haul transport optical networks such as the NSF network shown in Fig.~\ref{fig:topologyNSF} that covers the whole US mainland area. In order to investigate the effects of topology on the questions of interest, we also consider a symmetric network illustrated in Figure~\ref{fig:topologySymmetric} with the same number of nodes and network diameter as the NSF network (the link length is set to be 1330 km for this purpose) but with a different number of links and connectivity. The network must support a given traffic load as identified by a demand matrix specifying the source node,  destination node, and throughput requested of each traffic demand.

The network we investigate is fully elastic in that the spectrum assignment is flexible over a very fine grid. The central frequency and the bandwidth of each channel can both be flexibility tuned.  For mathematical simplicity, we model each channel as having a continuous-valued spectrum (i.e., not slotted) as an approximation to a fine grid.  In practice flexibility is achieved by using flexible-bandwidth transponders, which are the fundamental building blocks of EON. They transmit the signal on the fiber, performing all electrical-optical and optical-electrical conversions.  Without loss of generality, we assume a node structure with transponders based on orthogonal frequency division multiplexing (OFDM) technology. The configuration of each transponder  can be managed adaptively by the control plane as described in, for example, \cite{Casellas-2013}.

In this paper the resources being allocated to each demand include: connected fiber links forming a route, spectrum on each link, signal regeneration circuits on connecting nodes, and wavelength and/or modulation conversion at those nodes. (The MILP formulation given below could be easily modified to include other network attributes and constraints.)  The functionality that these resources provide is described in the following sections.

\subsection{Signal Regeneration}
When optical signals traverse the network, they often suffer from physical impairments such as signal loss, noise, dispersion and nonlinear effects. For networks with small physical dimensions (such as a local or metro area networks), the impairments can be ignored. In long-haul transport-scale networks, the impairments need to be accounted for in order to maintain acceptable signal quality (often referred to as the quality of transmission, QoT) at the receiving end. Due to the fact that impairments originate from many physical phenomena that accumulate over distance, and given that they usually depend on network state, for simplicity a conservative constraint on the length of fiber a signal can traverse before regeneration, called the transmission reach (TR), is often used to guarantee the QoT. If the source-destination distance on a route exceeds the TR, then regeneration is needed to reduce the physical impairments. The optical signal undergoes optical-electrical-optical (OEO) conversion at an intermediate node, and the regeneration (including re-amplifying, re-timing, re-shaping [3R]) is performed in the electrical domain. In our work, we assume 3R regeneration only occurs at intermediate nodes, not along the fiber links. To ensure proper QoT, the length of each transparent segment (part of the lightpath that has no intermediate regeneration) must be upper-bounded by the TR. As physical impairments depend on the bit rate and modulation scheme (spectral efficiency) used for a demand, so does the TR for the route used by that demand.

OEO conversion is expensive since it requires high-speed electronic equipment, and therefore regeneration needs to be carefully and conservatively planned.  In a typical optical network, not all nodes are equipped with regenerators to save on maintenance and other operational expenditures.

\subsection{Wavelength Conversion}
At a regeneration point, WC is often possible. We assume that the transponder can modulate the signal to an arbitrary new wavelength when converting the signal back from the electrical domain to the optical domain.

At nodes where no regenerators are available and therefore no WC is possible, the same wavelength must be used on both sides of the node on the lightpath.  This so-called wavelength continuity constraint causes spectrum fragmentation as small portions of the spectral resources become trapped between other connections with rigid wavelength assignments.
Due to asymmetry in topology and traffic, as network utilization increases certain links and nodes (shared by the most demands) may become bottlenecks. WC can be beneficial as a way of defragmenting the network by filling in gaps in the spectrum.  Defragmentation can lead to a significant reduction in total spectrum required by the system.

\subsection{Flexible Modulation}

Some networks may opt to invest in bandwidth-variable transponders that, in addition to being able to use a flexible number of OFDM subcarriers, can also modulate each subcarrier using a variety of different modulation schemes, such as binary phase-shift-keying (BPSK), quadrature-phase-shift-keying (QPSK), and high-order quadrature-amplitude-modulation (QAM). As the order of the modulation increases, so does the spectral efficiency of the transmission, requiring less bandwidth to transmit the same data-rate.

\begin{figure}[!tb]%[htdp]
   \centering
        \includegraphics[width=3.5in]{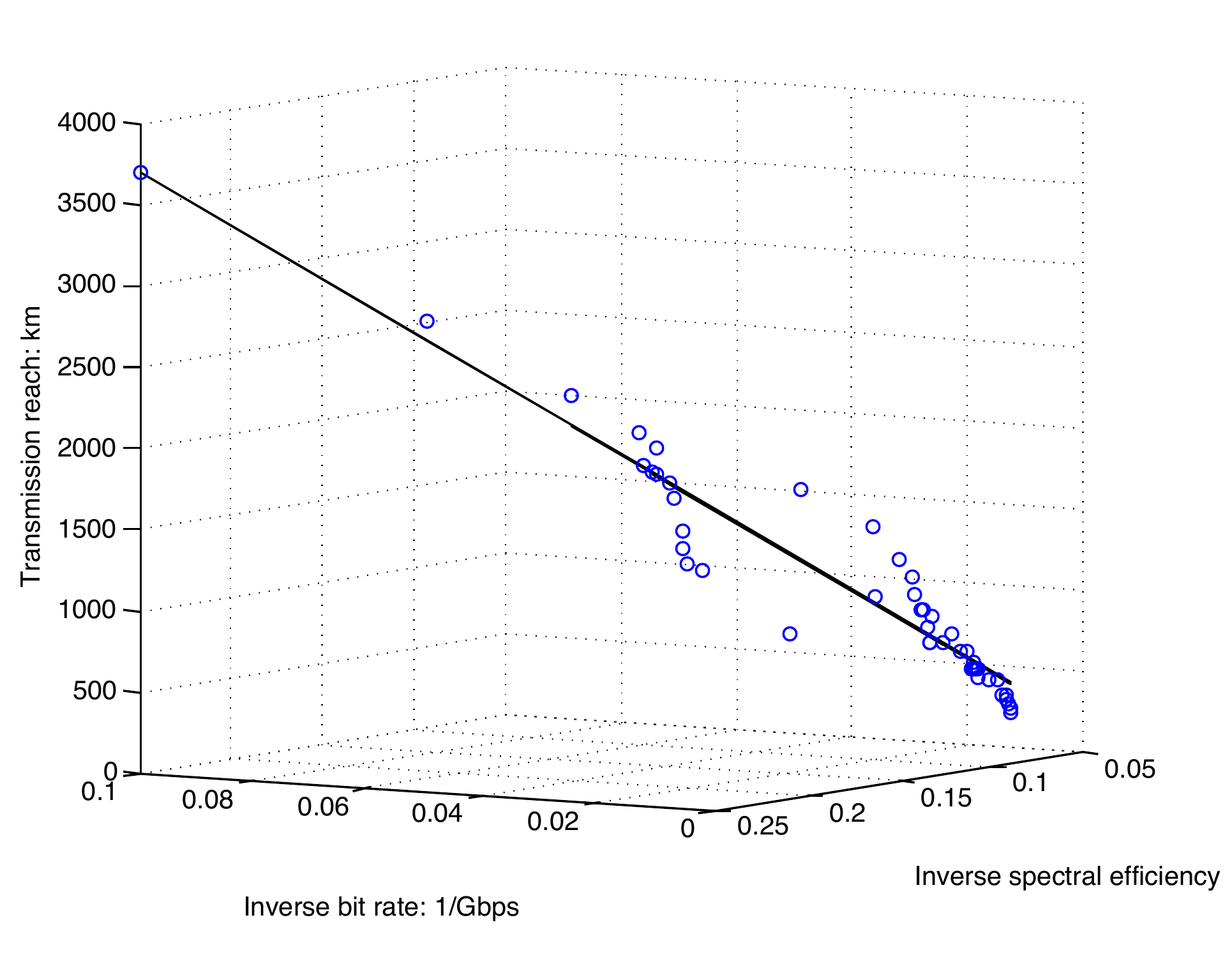}
    \caption{Transmission reach based on bit rate and spectral efficiency using polynomial fitting over experimental result data from \cite{Klekamp-2011}.}
    \label{fig:reach}
\end{figure}

The length of a transparent lightpath being used limits the spectral efficiency on that lightpath because signals with higher spectral efficiency are more susceptible to physical impairments.
By choosing an underlying modulation for a lightpath, one trades off data-rate and/or spectral efficiency for transmission distance.
In this paper we approximate the TR for a particular bit-rate $b$ and spectral efficiency $\eta$ by using published experimental data from \cite{Klekamp-2011}.  A linear regression model for the TR, denoted as $R$, results in the approximation shown in Figure~\ref{fig:reach}, corresponding to
\begin{equation}\label{eq:reach}
R(b,\eta) = \alpha b^{-1} + \beta \eta^{-1} + \gamma
\end{equation}
where, $\alpha, \beta, \gamma$ are coefficients optimized to fit the results in \cite{Klekamp-2011}. $R(b,\eta)$ is in unit of km, $b$ is in unit of Gbps, and $\eta$ is in unit of bit/symbol. The regression yielded values $\alpha=18600$, $\beta=8360,$ and $\gamma= -250$.

\subsection{Modulation Conversion}

%The spectral efficiency of traffic demands depends on the modulation scheme used, which in turn relates to the transmission reach of the demand based on its bit rate.
In EONs that use bandwidth-variable transponders that transmit demands at different data-rates and with multiple modulation schemes, the regenerators can also be allowed to modify the underlying modulation for each transparent segment. We refer to this functionality as modulation conversion (MC), which can occur only at a regeneration node.
If this functionality is not used, then only one modulation scheme is allowed for any single demand, and the spectral efficiency is then limited by the longest transparent segment. Since some transparent segments may be considerably shorter than others, those could have supported higher efficiencies. By allowing MC at regeneration nodes, the overall spectral efficiency can be improved and the required spectrum can be reduced.

%-----------------------------------------ILP formulations------------------------------------------------%
% 1. Design objective
% 2. Fundamental constraints
% 3. Constraints for each functions
% 4. Complexity estimation
\section{MILP} \label{sec:ilp}
% this is a general guideline of this section
In this section, we first introduce a basic link-based MILP formulation that solves a simple routing and spectrum assignment (RSA) problem, and then extend it to implement signal regeneration and multiple modulation schemes. Lastly we implement wavelength and modulation scheme conversion. Our general objective is to minimize the spectrum required by the system, as measured by the maximum frequency allocated over all links.  We also examine the impact of simultaneously optimizing the spectral use and regeneration resources using a multi-objective function.
%We show the complexity for each additional functionality over the general formulation.

The network is modeled as a graph {\bf$G(\mathcal{N,L})$} with $N$ nodes and $L$ uni-directional links.  We summarize the set notation used by our model in Table~\ref{tab:sets}.  The model also depends on parameters specific to the network configuration and the traffic demands. The notation for the independent parameters needed is given in Table~\ref{tab:parameters}.

\begin{table}[!t]
\caption{Sets used by Basic ILP}
\label{tab:sets}
\begin{tabular}{p{.34in}|p{2.8in}}
  \textbf{$\mathcal{N}$} & Set of nodes in the network.\\
  $\mathcal{L}$ & Set of unidirectional links in the network. Each link $L_{ij}$ is represented by its source and destination node, $L_{ij} \in \mathcal{L}$.\\
  ${\mathcal D}$ & Set of unidirectional traffic demands. Each demand $D_{sd}$ is represented by its source node $s$ and destination node $d$, $D_{sd} \in \mathcal{D}$.\\
\end{tabular}
\end{table}

\begin{table}[!t]
\caption{Parameters used by Basic ILP}
\label{tab:parameters}
\begin{tabular}{p{.34in}|p{2.8in}}
  $b_{sd}$ & Bit rate requested by demand $D_{sd}$.\\
  $\eta_{sd}$ &  Spectrum efficiency according to particular modulation scheme (e.g., 2 for QPSK).\\
   $S_{n,sd}$ & Relationship between nodes and demands: $S_{n,sd}=-1$ if node $n$ is the source node of demand $D_{sd}$ (i.e., $n = s$); $S_{n,sd}=1$ if node $n$ is the destination node of demand $D_{sd}$ (i.e., $n = d$); $S_{n,sd}=0$ otherwise (i.e., $n \neq s, n \neq d$).\\
  $G$ & Guard band in GHz.
\end{tabular}
\end{table}

The objective function of the MILP is to minimize the highest frequency required to support the network traffic:
\begin{equation}\label{eq:objective}
  \min_{F_{sd},V_{ij,sd},\delta_{sd,s'd'}} c,
 \end{equation}
where the optimization variables are defined in Table~\ref{table:variables}.
The optimization requires several constraints, listed below:

\begin{itemize}
\item Highest required spectrum:
\begin{eqnarray}\label{eq_c}
  c \geq F_{sd} + B_{sd} \quad \forall D_{sd}\in \mathcal{D},
\end{eqnarray}
where $B_{sd}$ is the bandwidth required by $D_{sd}$ for a given $\eta_{sd}$,  $B_{sd} = b_{sd} \times \eta_{sd}^{-1}$.

\item Flow conservation constraints:
\begin{align}
  \sum_{L_{ij}\in \mathcal{L}, j=n} V_{ij,sd} - \sum_{L_{ij}\in \mathcal{L}, i=n} V_{ij,sd} \;=\; S_{n,sd} \nonumber \\ \forall n\in \mathcal{N},\,  D_{sd}\in \mathcal{D}
  \end{align}

\item No spectrum overlap constraints, $\forall D_{sd},D_{s'd'} \in \mathcal{D}$ :
\begin{align}
  \delta_{sd,s'd'} \! + \! \delta_{s'd',sd}  \; = \;  & 1 \label{eq_noover_1} \\
  F_{sd} \!  -  \! F_{s'd'} \; \leq \; & T (1 \! - \! \delta_{sd,s'd'} \!  + \!  2  \! -  \! V_{ij,sd} \!  -  \! V_{ij,s'd'}) \label{eq_noover_2} \\
F_{sd} \!  -  \! F_{s'd'} \!   + \!  B_{sd} \!  + \!  G \; \leq \; & (T \! + \! G) \nonumber \\ & \times  (1 \! - \! \delta_{sd,s'd'}  \! + \!  2 \!  -  \! V_{ij,sd} \!  - \!  V_{ij,s'd'}) \label{eq_noover_3}
\end{align}
where $T$  is the total spectrum required by the network traffic, $T = \sum_{D_{sd} \in \mathcal{D}} b_{sd} \times \eta_{sd}^{-1}$.
\end{itemize}

Eqs. (\ref{eq:objective}) to (\ref{eq_noover_3}) define a general link-based RSA formulation for EON.  Together Eqs. (\ref{eq_noover_1})-(\ref{eq_noover_3}) enforce a contiguous spectrum assignment to each demand. Eq.~(\ref{eq_noover_1}) says that for any two demands $sd$ and $s'd'$ that share a link, one demand has to have a starting frequency lower than the other, and therefore one of the ordering variables is zero and the other is one. Eq.~(\ref{eq_noover_2}) enforces the necessary relationship between starting frequencies of the two demands based on the variable $\delta_{sd,s'd'}$. Then Eq.~(\ref{eq_noover_3}) forces the starting frequency of the demand with the higher starting frequency to be far enough away from the starting frequency of the lower adjacent channel, i.e., provides room for the signal bandwidth and guard band. These expressions can be modified to implement the more sophisticated signal processing we  consider in this paper.  Each functionality is discussed below, together with the additional variables and constraints needed.

\begin{table}[!t]
\caption{Variables used by Basic ILP}
\label{table:variables}
\begin{tabular}{p{.34in}|p{2.8in}}
  $F_{sd}$ & Starting frequency index of demand $D_{sd}$.\\
  $V_{ij,sd}$ & Link assignment: $V_{ij,sd}=1$ if link $L_{ij}$ is assigned to demand $D_{sd}$; $V_{ij,sd}=0$ otherwise.\\
  $\delta_{sd,s'd'}$ & Order of the starting frequency index of demand $D_{sd}$ and $D_{s'd'}$.\footnotemark $\delta_{sd,s'd'}=1$ if $F_{sd}\le F_{s'd'}$,  $\delta_{sd,s'd'}=0$ if $F_{sd} > F_{s'd'}$.\\
$c$ & Highest frequency index required by the network traffic.\\
\end{tabular}
\end{table}
\footnotetext{This relationship between two demands is only of interest if they share a link. We use this relationship in the following constraints to guarantee no overlapping between spectra assigned to multiple demands.}

\subsection{Multiple modulation schemes}
When each demand has different spectral efficiency, their transmission reach also varies. This is implemented by making $\eta_{sd}^{-1}$, the inverse spectral efficiency of demand $D_{sd}$, a variable instead of a constant parameter. In our model we relax this value from its normal discrete nature to be a real number bounded by the largest and smallest inverse spectral efficiencies allowed: $\eta_{sd,{\text{MIN}}}^{-1} \leq \eta_{sd}^{-1} \leq \eta_{sd,{\text{MAX}}}^{-1}$.

\subsection{Signal regeneration}

Signal regeneration can be used to increase the length of a lightpath beyond the transmission reach. The following constraints, using additional parameters and variables defined in Tables~\ref{table:parameters_tr} and \ref{table:variables_tr}, respectively, must be satisfied so that the QoT requirements are fulfilled for all demands.
\begin{table}[!t]
\caption{Parameters Used by transmission reach constraint}
\label{table:parameters_tr}
\begin{tabular}{p{.34in}|p{2.8in}}
 $\ell_{ij}$ &Length of link $L_{ij}$ in km.\\
 $R_{sd}$ & Transmission reach of demand $D_{sd}$ according to particular spectral efficiency, e.g., in Eq.~(\ref{eq:reach})\\
 $\mathcal{N}^r$ & Set of regeneration nodes. \\
\end{tabular}
\end{table}

\begin{table}[!t]
\caption{Variables used by transmission reach constraint}
\label{table:variables_tr}
\begin{tabular}{p{.34in}|p{2.8in}}
$Y_{n, sd}$ & $Y_{n, sd}=0$ if node $n$ is not on the lightpath assigned to demand $D_{sd}$. Otherwise, $Y_{n, sd}$ is the physical distance from node $n$ on the lightpath to the beginning of that transparent segment for demand $D_{sd}$.\\	
$U_{ij, sd}$ & $U_{ij, sd}=0$ if the entire link $L_{ij}$ is not assigned to demand $D_{sd}$ ($V_{ij,sd}=0$). Otherwise, $U_{ij, sd}$ is the physical distance from node $i$ to the beginning of the transparent segment for demand $D_{sd}$. Equivalently, if we were not restricted to linear functions, we could have defined $U_{ij, sd}=V_{ij,sd} Y_{i,sd}$.\\
\end{tabular}
\end{table}

We consider two cases. In the first case, the nodes in the network that are equipped with regeneration circuits have been pre-selected.  There has been quite some research recently on how to select regeneration nodes, including \cite{Chen-2012}. The constraints that the MILP must satisfy for all $D_{sd} \in \mathcal{D}$ and $ L_{ij} \in \mathcal{L}$ are as follows:
\begin{align}
  U_{ij,sd} &\;\leq \; V_{ij,sd} R \label{eq:trc1}\\
  U_{ij,sd} &\; \leq  \; Y_{i,sd} \label{eq:trc2} \\
  Y_{i,sd} - U_{ij, sd} &\; \leq\;  R  (1 - V_{ij,ld})\label{eq:trc3}
  %Y_{n,sd} & \; = \; 0, \quad \forall n\in \mathcal{N}^{r} \mbox{ or }n=s \label{eq:reg1}
    \end{align}
%\begin{equation}\label{eq:reg2}
%  Y_{n,sd}  = \!\!\!\!\!  \sum_{L_{ij}\in \mathcal{L}:j=n} \!\!\!\!\! U_{ij,sd} + \ell_{ij} V_{ij,sd} \\\quad \forall n\notin \mathcal{N}^r \mbox{ and }n \neq s
%  \end{equation}

\begin{equation}\label{eq:reg2}
Y_{n,sd} = \left \{
  \begin{array}{l l}
    \sum_{L_{ij}\in \mathcal{L}:j=n} \!\!\! \ell_{ij} V_{ij,sd} & \text{if }\exists L_{ij}, i \in \mathcal{N}^r \\  & \mbox{ and } V_{ij,sd} = 1\\
    \sum_{L_{ij}\in \mathcal{L}:j=n} \!\!\! U_{ij,sd} + \ell_{ij} V_{ij,sd} &  \text{otherwise}
  \end{array} \right.
 \end{equation}
where $R=R(b_{sd},\eta_{sd})$ from Eq.~(\ref{eq:reach}). For cases when node $i$ is an intermediate node but link $L_{ij}$ is not an intermediate link for demand $D_{sd}$ (i.e., $V_{ij,sd} = 0$), the distance $U_{ij,sd}=0$ but $Y_{i,sd}$ is not necessarily zero but has to be less than the transmission reach. In this case Eq.~(\ref{eq:trc3}) reduces to $Y_{i,sd} \leq R$.

The second case we consider is one where the regeneration nodes are not pre-selected.  We then use the MILP to optimize the placement of regeneration equipment on the network. We treat the regeneration node assignments as binary variables $I_n$ and, with the help of additional variables defined in Table~\ref{table:variables_regen}, we use Eqs. (\ref{eq:trc1})-(\ref{eq:trc3}) from above and replace Eq.~(\ref{eq:reg2}) by % (\ref{eq:reg1}) and (\ref{eq:reg3}) and
\begin{align}
  %Y_{n,sd} \; &\leq \; 0\label{eq:reg3}\\
  Y_{n,sd} \; & = \; \sum_{L_{ij}\in \mathcal{L}:j=n} \!\!\!\!\! X_{ij,sd} + \ell_{ij}  V_{ij,sd}, \label{eq:reg4}
\end{align}

\begin{figure}[!tb]%[htdp]
	\centering\includegraphics[width=2in]{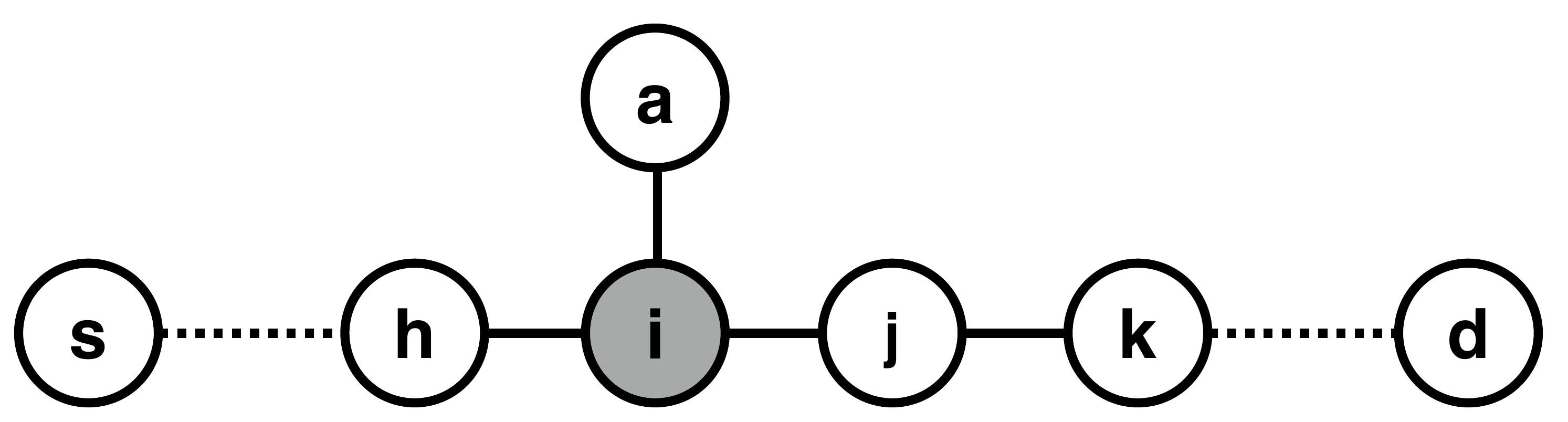}
	\caption{Illustration used to explain the variables defined for constraining the transmission reach}\label{fig:TRillustrate}
\end{figure}

Consider Fig. \ref{fig:TRillustrate} to understand the TR constraints. There is a demand $D_{sd}$ and a path from node $s$ to node $d$, while link $L_{ia}$ does not belong to the path of demand $D_{sd}$. Assume node $i$ is the only regeneration node on the path. Inequality (\ref{eq:trc1}) says that for any link on the path (e.g., $L_{ij}$), $U_{ij,sd} \le R$. Since node $i$ is the only regeneration node, $U_{ij}$ for link $L_{ij}$ and  $Y_{i,sd}$ both represent the distance from node $s$ to node $i$. Since link $L_{ia}$ is not on the path for demand $D_{sd}$ (i.e., $V_{ia,sd} = 0$), $U_{ia, sd} = 0$.
Eq.~(\ref{eq:reg2}) says that, for node $i$, since no link that leads to it starts with a regeneration node, $Y_{i,sd}$ is the sum of $U$'s for all links that lead to node $i$ plus the link length of the link that is on the path. Since all links except for link $L_{hi}$ have $U$ equal to zero, $Y_{i,sd}$ is the sum of $U_{hi,sd}$ and link length of link $L_{hi}$. But, for node $j$, since node $i$ is a regeneration node, $Y_{j,sd}$ is just the link length of link $L_{ij}$. Inequalities (\ref{eq:trc2}) and (\ref{eq:trc3}) say that $Y_{i,sd} < R$, $U_{ia,sd} < Y_{i,sd}$, and $U_{ij,sd} = Y_{i,sd}$. When the allocation of regeneration nodes is unknown, we use the variable $X_{ij,sd}$ to differentiate the cases when node $i$ is a regeneration node or not, as in Eq.~(\ref{eq:reg4}).

\begin{table}[!t]
\caption{Variables used for Regenerator Circuit Assignment}
\label{table:variables_regen}
\begin{tabular}{p{.34in}|p{2.8in}}
$I_n$ & Regeneration nodes: $I_n = 1$ if node $n$ is used as a regeneration node; $I_n=0$ otherwise.\\
$N_{n,c}$ & Number of regeneration circuits used on node $n$. \\
$I_{n,sd}$ & Regeneration at node $n$: $I_{n,sd} = 1$ if demand $D_{sd}$ is regenerated at node $n$, $I_{n,sd} = 0$ otherwise. For $I_{n,sd} = 1$, node $n$ has to be a regeneration node, i.e., $I_n = 1$, but also its regeneration circuit has to be used by demand $D_{sd}$.\\
$X_{ij,sd}$ & Distance used to calculate $Y_{n,sd}$ based on whether regeneration occurs at node $i$:  $X_{ij,sd} = U_{ij,sd}$ if $I_{i,sd} = 0$, $X_{ij,sd} = 0$, otherwise.\\
\end{tabular}
\end{table}

Constraints that limit the number of OEO circuits per regeneration node can also be included using:
\begin{align}\label{rnc}
  N_{n,c} \;  & =  \; \sum_{D_{sd} \in \mathcal{D}} I_{n,sd} \\
  I_n N_{n,c\text{MAX}} \; &  \geq \; N_{n,c}
 \end{align}
where $N_{n,c\text{MAX}}$ is the largest number of regeneration circuits that can be equipped on a regeneration node.

When the cost of regeneration resources is a concern, we can build a multi-objective function to balance the cost of regeneration and spectrum resources:
\begin{eqnarray}\label{eq:multiple}
  \min \left\{ ac + (1-a) \sum_{n \in \mathcal{N}} I_n  \right\}
  \end{eqnarray}
where the coefficient $a \in [0, 1]$  represents the cost relationship between using the two resources. This objective function minimizes the total cost of all resources together, according to their relative costs. While we do not presume to know the exact cost relationship among the two, a network designer can base their objective function on realistic requirement, and use Eq.~(\ref{eq:multiple}) to determine what resources are needed and where.

%%****wavelength conversion is only performed at regeneration node, and regeneration is also done at intermediate node (not mid-link due to that 3R requires OEO, I think we talk about this in section Network description. So I wonder if we need to explain again here.

\subsection{Wavelength and Modulation Conversion}
%Assuming there are finite modulation schemes available and each modulation scheme corresponds to particular spectral efficiency. Modulation schemes together with the bit rate requirement of the traffic demands determine the demands' transmission reach. The path-based MILP formulation shown in \cite{Christodoulopoulos-2013} considers the transmission reach constraint by listing a fixed number of route and modulation pairs for each demand. Regeneration nodes are assigned to each route and modulation pair in a 'when necessary' fashion. Those pairs the route in which has link too long for the modulation and bit rate of the demand is then excluded from the list. The rest of the list is then used to solve the problem. Such algorithm lacks the flexibility of assigning regeneration nodes in that it minimizes the number of regeneration nodes for each route and modulation pair but does not guarantee such assignment is optimal for the whole network. We therefore propose the link-based formulation where all potential route and modulation and regeneration node assignment are considered

\begin{table}[!t]
\caption{Variables used by Wavelength and Modulation Conversion}
\label{table:variables_wc}
\begin{tabular}{p{.34in}|p{2.8in}}
$F_{ij,sd}$ & Starting frequency index of demand $D_{sd}$ on link $L_{ij}$.\\
$\eta_{sd,ij}^{-1}$ & Inverse spectral efficiency of demand $D_{sd}$ on link $L_{ij}$.\\
\end{tabular}
\end{table}

When WC is available, the frequencies used for a demand can be different on the links entering a regeneration node and exiting it. In order to represent this flexibility we define the starting frequency on a link-by-link basis, as shown in Table~\ref{table:variables_wc}, and re-write Eqs. (\ref{eq_noover_2}) and (\ref{eq_noover_3}), the constraint that guarantees no spectrum overlap, for all $ n\in \mathcal{N}$ as:
 \begin{align}\label{str:wav}
   \sum_{L_{ij} \in \mathcal{L}:j = n}\!\!\!\!\!\! F_{ij,sd} -\!\!\!\!\!\! \sum_{L_{ij} \in \mathcal{L}: i = n}\!\!\!\!\!\!  F_{ij,sd} & \; \geq \;  -T \times (I_{n,sd} + |S_{n,sd}|) \nonumber  \\
   \sum_{L_{ij} \in \mathcal{L}:j = n}\!\!\!\!\!\!  F_{ij,sd} - \!\!\!\!\!\! \sum_{L_{ij} \in \mathcal{L}: i = n} \!\!\!\!\!\! F_{ij,sd} & \; \leq \;  T \times (I_{n,sd} + |S_{n,sd}|)
\end{align}
This constraint requires that if node $n$ is an intermediate node for demand $D_{sd}$, i.e., $n \neq s, n\neq d$ and $n$ is not used as a regeneration node for demand $D_{sd}$, then the starting frequency assignments entering node $n$ equals the starting frequency assignments exiting node $n$. For other cases, this constraint does not apply.

Similar to WC, when MC is available the spectral efficiency of each demand on each link can be different than its immediate uplink or downlink if the joining node is used as a regeneration node. We must define the spectral efficiency on a link-by-link basis, as listed in Table~\ref{table:variables_wc}.
The MC constraint can be written as:
 \begin{align}\label{str:mod}
 \sum_{L_{ij} \in \mathcal{L}:j = n}\!\!\!\!\!\!\!\!  \eta_{sd,ij}^{-1} - \!\!\!\!\!\!\sum_{L_{ij} \in \mathcal{L}: i = n}\!\!\!\!\!\!\!\! \eta_{sd,ij}^{-1} & \;  \geq \; -\eta_{sd, \text{MAX}}^{-1} \times (I_{n,sd} + |S_{n,sd}|) \nonumber  \\
   \sum_{L_{ij} \in \mathcal{L}:j = n} \!\!\!\!\!\!\! \! \eta_{sd,ij}^{-1} - \!\!\!\!\!\!\sum_{L_{ij} \in \mathcal{L}: i = n}\!\!\!\!\!\!\!\! \eta_{sd,ij}^{-1} & \; \leq  \; \eta_{sd, \text{MAX}}^{-1} \times (I_{n,sd} + |S_{n,sd}|) \nonumber \\
   \end{align}
Similar to the WC constraint, this constraint requires that modulation (equivalently, spectral efficiency) only be converted at nodes $n$ where the demand is regenerated, i.e., where $I_{n,sd} = 1$.

\section{Recursive MILP}\label{sec:recursive}
%-----------------------------------------Recursive ILP solutions---------------------------------------%
% 1. Description
% 1.1 Motivation
% 2. Optimality estimation
% 3. Complexity estimation
The computational complexity of the MILP formulation for EON restricts its implementation to offline calculation only. And even then, it does not scale well as the size of the network or the number of traffic demands increase. An effective alternative would be to reduce the problem size to such an extent that the results can be found in acceptable time and with reasonable computational resources. Many works have attempted to break the whole RSA problem into sub-problems with or without losing some degree of optimality \cite{Wan-2012}, \cite{Liu-2013c}. In this section, we propose a different way to reduce the problem size by splitting the traffic matrix into sub-matrices and solving them sequentially, a technique we call {\em recursive MILP}.

The recursive MILP approach is motivated by the understanding that the complexity of the problem is greatly affected by the number of traffic demands that need to be accommodated at once. By separating them into subsets and allocating those in sequential iterations, the overall runtime as well as other computational resources, such as memory, can be reduced. The solution from the previous iteration forms new MILP constraints for the new iteration. In particular, the first iteration can be viewed as a subproblem with the same constraints but with fewer demands. In the new iterations, the constraints (e.g., non-overlapped spectrum assignment) apply to both the assigned and unassigned resources. In this manner, the original problem can be solved, albeit not optimally, after all iterations are done.

Another advantage of using recursive MILP is that the complexity is easy to estimate. For example, in the aforementioned MILP formulation, the complexity-dominating variable is $\delta_{sd,s'd'}$, which grows with the number of demands $|{\mathcal D}|$ squared, i.e., $O(|{\mathcal D}|^2)$. By solving the same problem recursively, the number of variables of each calculation is reduced. If the number of subsets is $S$, the last subproblem (which has the highest number of variables) will have about $\frac{|{\mathcal D}|^2}{S^2}$ many $\delta_{sd,s'd'}$. Running the MILP in recursive mode does not require reformulating the problem: the constraints remain the same but the variables that represent demands from previous iterations become constants.

It can be expected that the recursive MILP suffers loss of optimality compared to the non-recursive counterpart. The gap between the sub-optimal solution from the recursive MILP and the optimal solution depends on the size of the subset and the grouping and ordering of traffic demands. Since the complexity is easy to estimate, network designers can base the implementation of the formulation on the complexity they can accept.
In Table~\ref{tab:recursive} we show a comparison between the order of computational complexity for non-recursive (single-run) and recursive MILP solutions.
%, using the symbol $I$ as the number of iterations.

\begin{table*}[t]
\caption{Complexity of One Iteration of Recursive and Non-recursive Basic MILP}\label{tab:recursive}
\begin{center}
{\small \begin{tabular}{|c|c|c|c|}
         %\hline  \multicolumn{3}{|c|}{Entry}  \\
         \hline \multicolumn{2}{|c|}{Number of Variables}    &  \multicolumn{2}{|c|}{Number of Constraints} \\
         \hline
          non-recursive & recursive & non-recursive & recursive \\
         \hline  $(1 + L+ |{\mathcal D}|) \times |{\mathcal D}| + 1$   & $(1+ L+|{\mathcal D}|/S) \times |{\mathcal D}|/S $ &  $|{\mathcal D}| + N|{\mathcal D}| + (1 + 2L) \times |{\mathcal D}|^2$ & $|{\mathcal D}|/S + N|{\mathcal D}|/S  + (1 + 2L) \times (|{\mathcal D}|/S) ^2$ \\
         \hline
\end{tabular}}
\end{center}
\end{table*}

The complexity and optimality are affected not just by the size of the demand subsets, but also the selection of demands in each iteration. The grouping of demands can be done in many ways such as sorting them randomly or based on their characteristics such as volume, locality etc. In Section~\ref{sec:simulation} we show a comparison of the required spectrum using different ordering schemes.

Another use for the recursive MILP implementation is to help accommodate network expansion with existing infrastructure. The existing assignments of physical resources (links, regeneration devices, spectrum) can be input into the MILP as constraints, in the same way as was done for results from earlier iterations of the recursive MILP. In our simulation results below we show a progressively increasing number of traffic demands, each demand set including the same demands from previous sets. The spectrum predicted shows the resources required for network expansion.  This problem is  similar to the so-called {\em dynamic} resource allocation problem \cite{Zhu-2013}, \cite{Kuang-2014}, the difference being whether connections are also torn-down. In that scenario, the above MILP can also be used is a similar recursive way as was shown for WDM systems in \cite{Wang-2012b}.

\section{Numerical Results}\label{sec:simulation}

We test our MILP formulations on the NSF-24 mesh network, shown in Fig.~\ref{fig:topologyNSF}, which is often used as a benchmarking topology in the literature.  In addition, to gauge the sensitivity of our results on network topology we simulate a symmetric network illustrated in Fig.~\ref{fig:topologySymmetric}, with 24 edge nodes that also serve as intermediate nodes for routing traffic. The NSF-24 network has 43 by-directional links (or uni-directional link pairs of 86 links), while the symmetric-24 network has 55 by-directional links (or 110 uni-directional links). In our simulation, we assume traffic demands are generated between random selected node pairs and have random bit-rate requests ranging uniformly from 1 to 100 Gbps in order to represent the heterogeneity of Internet traffic. The traffic is assumed static, and no traffic grooming or reverse grooming (i.e., traffic splitting) is considered. In all cases we collect simulation results over 20 independent random demand sets and report average results.

%\begin{figure}[htb]%[htdp]
%\vspace{-.1in}
%   \centering
%        \includegraphics[width=\columnwidth]{placeholder}
%    \caption{This is a place holder plot}
%\vspace{-.1in}
%    \label{fig:placeholder}
%\end{figure}

In the numerical results, we investigate one or two features at a time in each section.  Unless otherwise stated, the algorithm tested is an optimum MILP (non-recursive) assuming multiple modulations are available with $\eta_{sd} \in [1,10]$ and signal regeneration capability at all nodes, but no modulation or wavelength conversion is used so that each demand uses the same modulation and spectrum from source to destination.

\subsection{Recursive and Non-recursive MILP}
\begin{figure}[!tb]%[htdp]
 \centering       \includegraphics[width=3.3in]{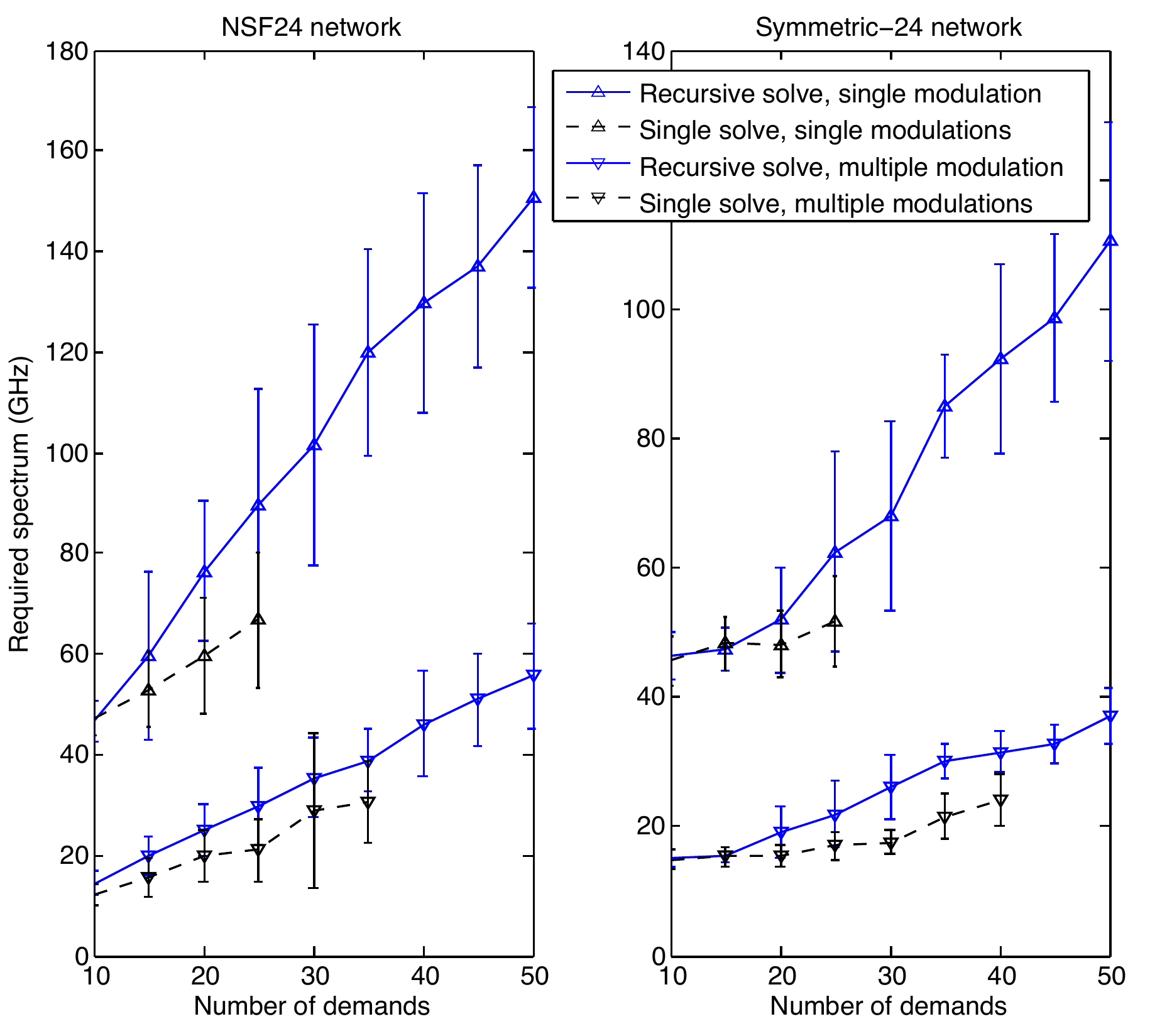}
    \caption{Required spectrum using the recursive MILP and single solve MILP for a single modulation scheme ($\eta = 2$) and multiple modulation schemes ($1 \le \eta \le 10$)}
    \label{fig:recursive}
\end{figure}

We first verify the applicability of the MILP solution to the network sizes we have chosen, and compare the optimality and computational complexity between the recursive and non-recursive approaches.  In Fig.~\ref{fig:recursive} we show the required spectrum for both single and multiple modulation schemes. We also plot the standard deviation for  our results. As the number of demands allocated increases, the spectral usage increases approximately linearly. For the single modulation case, we choose (arbitrarily) $\eta_{sd}=2\,\,\forall D_{sd}$ (QPSK), while for the multiple modulation case $\eta_{sd}$ is optimized.  When the MILP assigns resources to all demands together (the ``single solve'' approach), the performance is notably better than the recursive approach (assuming a random partition of the demands into sets of 5), but too computationally burdensome for more than 30-40 simultaneous traffic demands.  We also conclude that the added flexibility of optimizing the spectral efficiency for each demand more than halves the required spectrum.  Both networks show similar results.

%I removed this: so that all demands can reach their destinations with regeneration

\begin{figure}[!tb]%[htdp]
   \centering
        \includegraphics[width=3.3in]{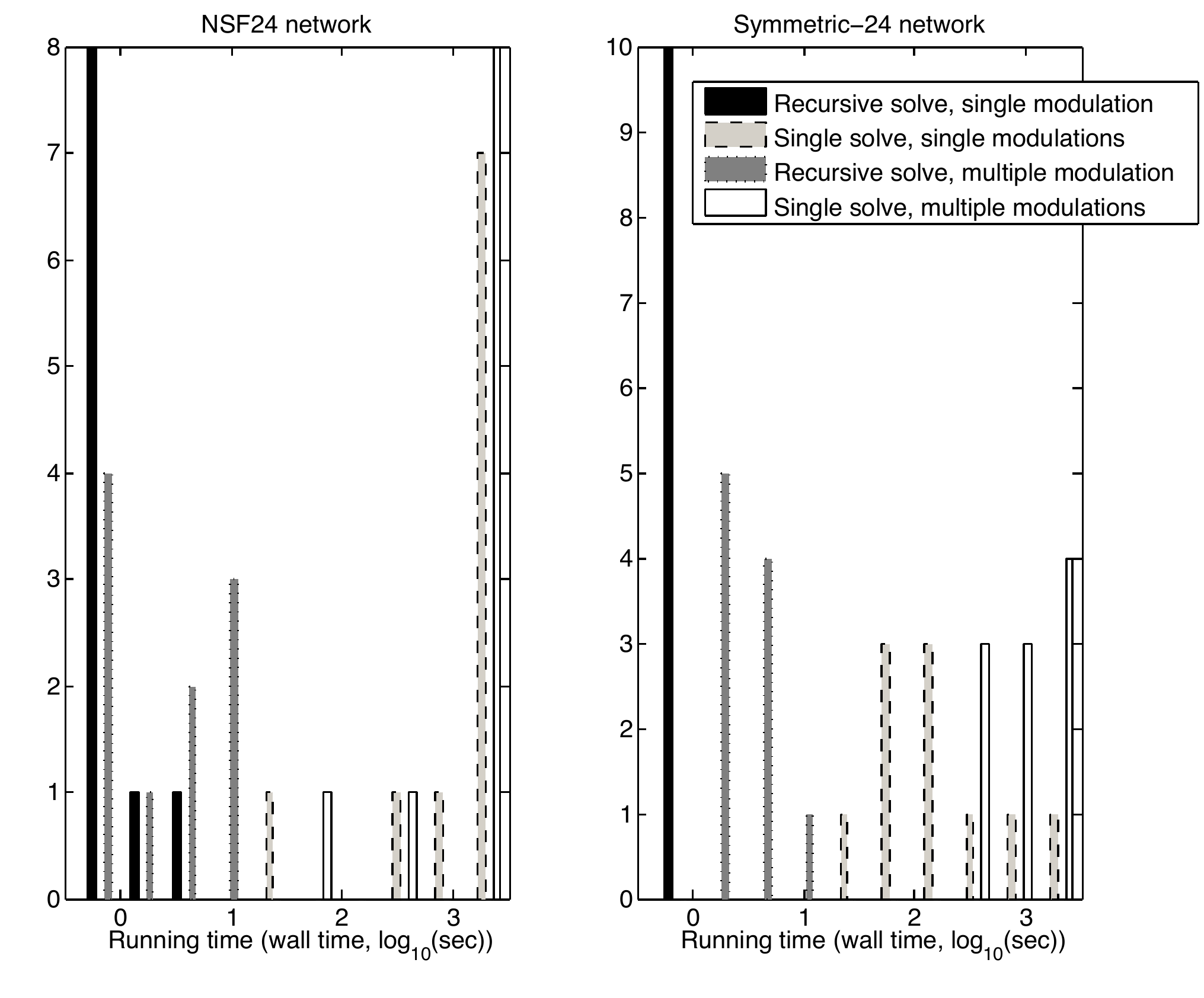}
    \caption{Running time for the recursive MILP and single solve MILP for a single modulation scheme ($\eta = 2$) and multiple modulation schemes ($1 \le \eta \le 10$)}
    \label{fig:recursiveruntime}
\end{figure}

In Fig. \ref{fig:recursiveruntime} we compare the complexity of the four approaches from Fig.~\ref{fig:recursive}. We show a histogram of computation running times when the total number of demands is 25. The running times for the optimal MILP vary considerably between trials (we set a time limit of 3000 seconds), while the running times for the recursive approach  are uniformly short.

\begin{figure}[!tb]%[htdp]
   \centering
        \includegraphics[width=3.3in]{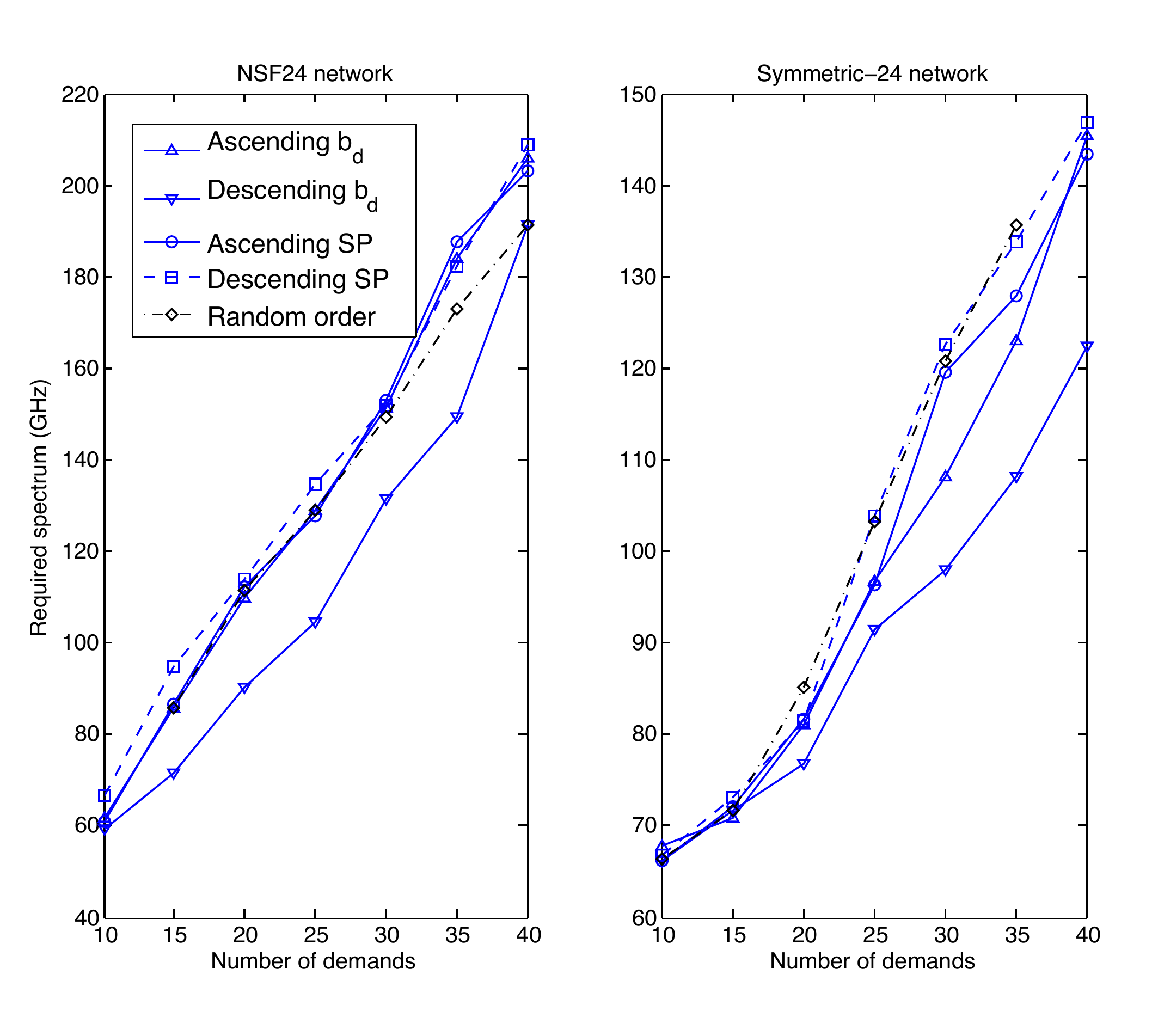}
    \caption{Required spectrum for the recursive MILP with different ordering schemes for the same 40 demands, as the resource assignment for the demand subsets progresses. }
    \label{fig:recursiveorder}
\end{figure}

In Fig \ref{fig:recursiveorder} we show that the partitioning and ordering of traffic demands in the recursive solution has a small but non-negligible impact on the required spectrum. When only a few demands have been assigned resources, accommodating high data-rate demands first leads to a lower required spectrum. Also, assigning demands that have the shortest shortest-paths (labeled ``SP'') first typically requires less spectrum than assigning the longer-distance connections first.  The differences are slightly more pronounced on the symmetric network.

%\begin{figure}[!htb]%[htdp]
%\vspace{-.1in}
%   \centering
%        \includegraphics[width=\columnwidth]{p2j_recursive_mm_c}
%    \caption{Required spectrum with multiple modulation scheme for recursive solve and single solve}
%\vspace{-.1in}
%    \label{fig:recursive_mm}
%\end{figure}

\begin{figure}[!tb]%[htdp]
   \centering
        \includegraphics[width=3.3in]{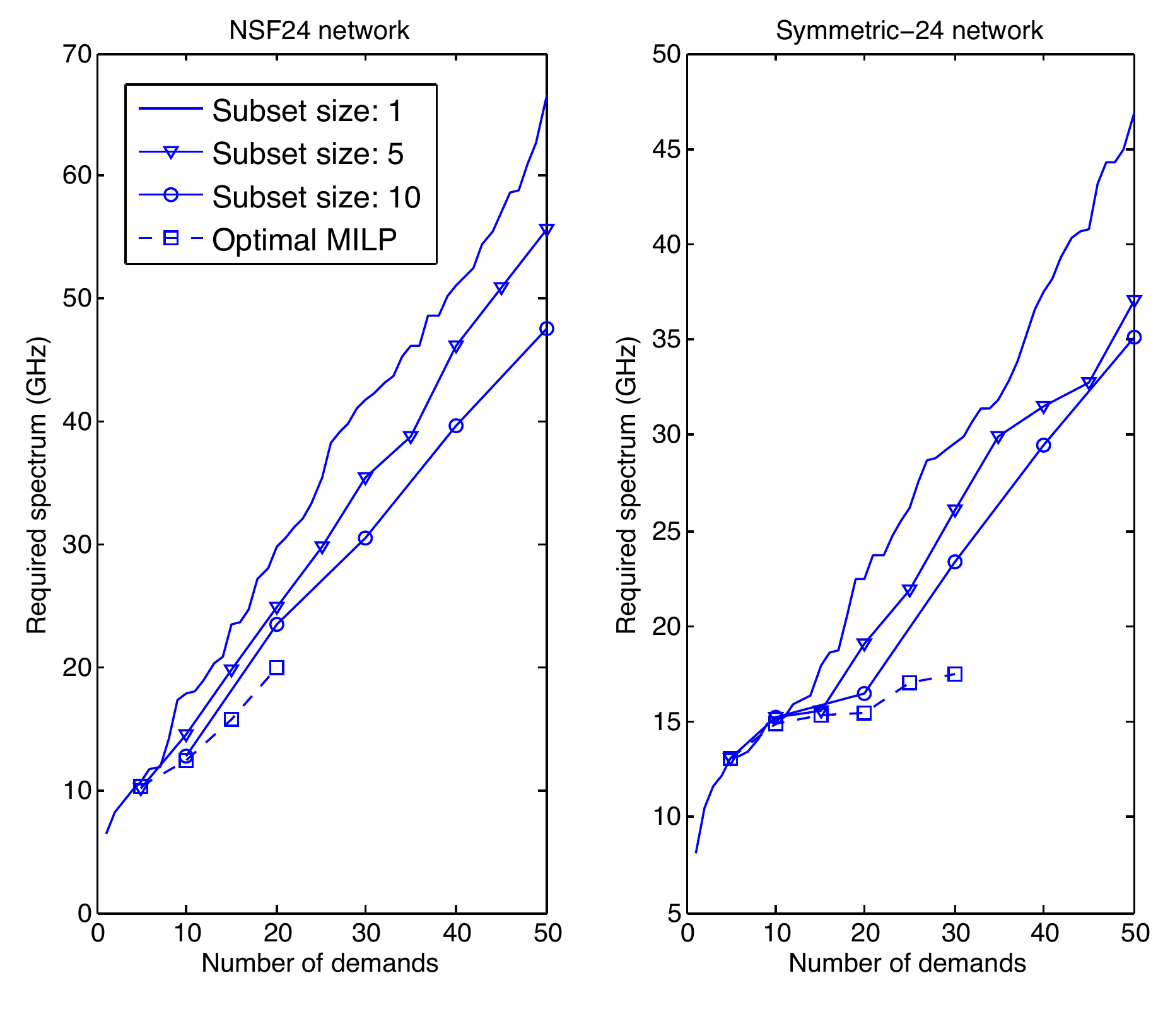}
    \caption{Required spectrum by solving the same recursive formulation with different demand subset sizes}
    \label{fig:step}
\end{figure}

\begin{figure}[!tb]%[htdp]
   \centering
        \includegraphics[width=3.3in]{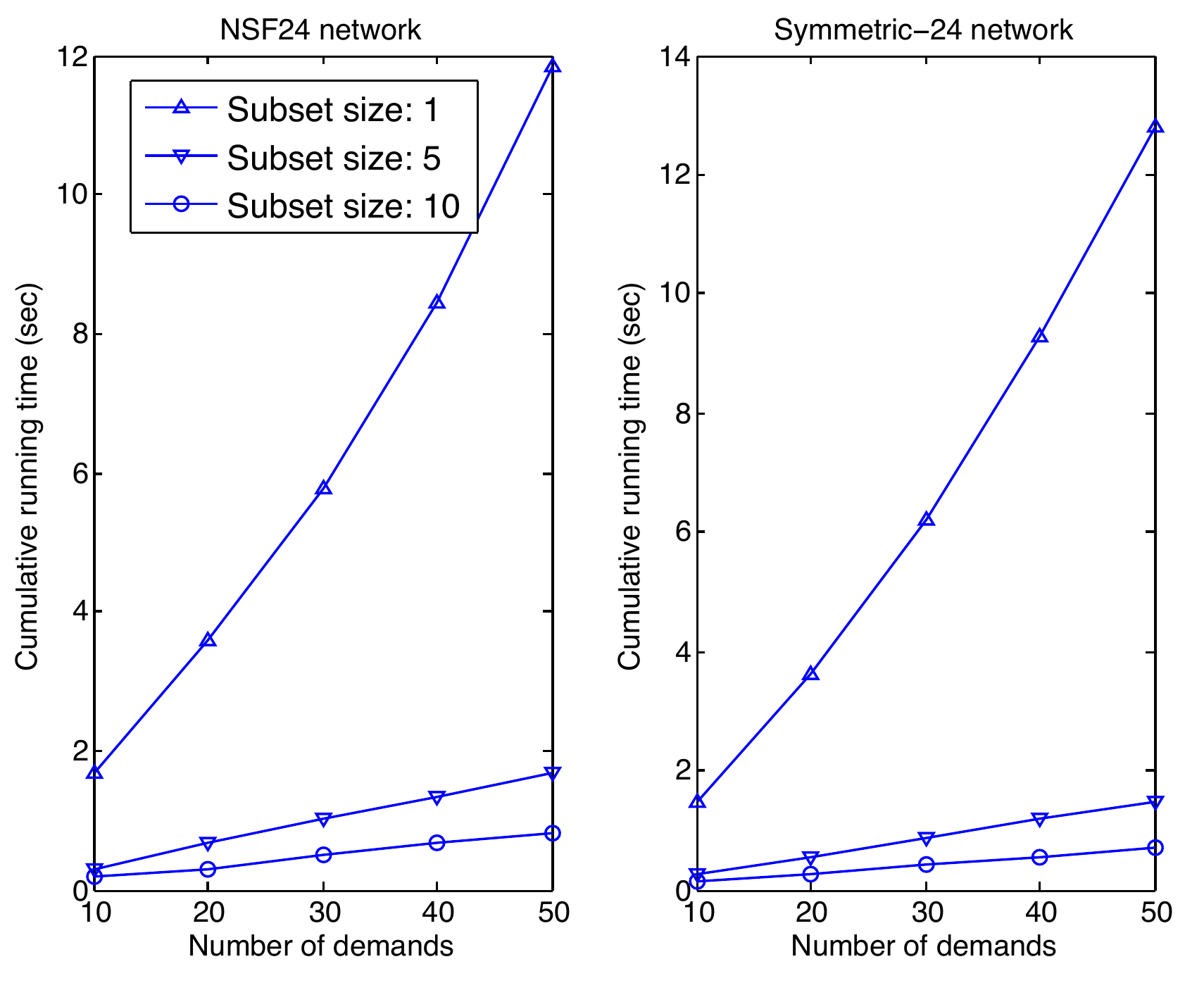}
    \caption{Cumulative running time by solving the same recursive formulation with different demand subset sizes}
    \label{fig:stepruntime}
\end{figure}

We then investigate the effect of the demand subset size $|{\mathcal D}|/S$ on the required spectrum obtained by the recursive MILP. The single solve approach, which finds the globally optimal result for all the traffic demands together, must always yield the smallest required spectrum. In Fig.~\ref{fig:step}, we see that as the subset size decreases, the required spectrum increases. When few demands are assigned per iteration, the required spectrum appears to be stair-stepped, since new connections can often use gaps in the allocated spectrum left by fragmentation induced by the sub-optimal resource allocation.  In Fig.~\ref{fig:stepruntime}, we show the cumulative running time that each case needs to assign a certain number of demands. Having a smaller subset size requires more time to accommodate traffic. As more demands are assigned, the new demands also require more time to be assigned because the solutions from previous subsets have to be used as constraints on the existing network state.
%due to assigning new spectrum to a few addition demands, which then takes a few more demands to fully re-use the spectrum throughout the network.

\subsection{Multi-objective Formulation}
\begin{figure}[!tb]%[htdp]
   \centering
        \includegraphics[width=3.3in]{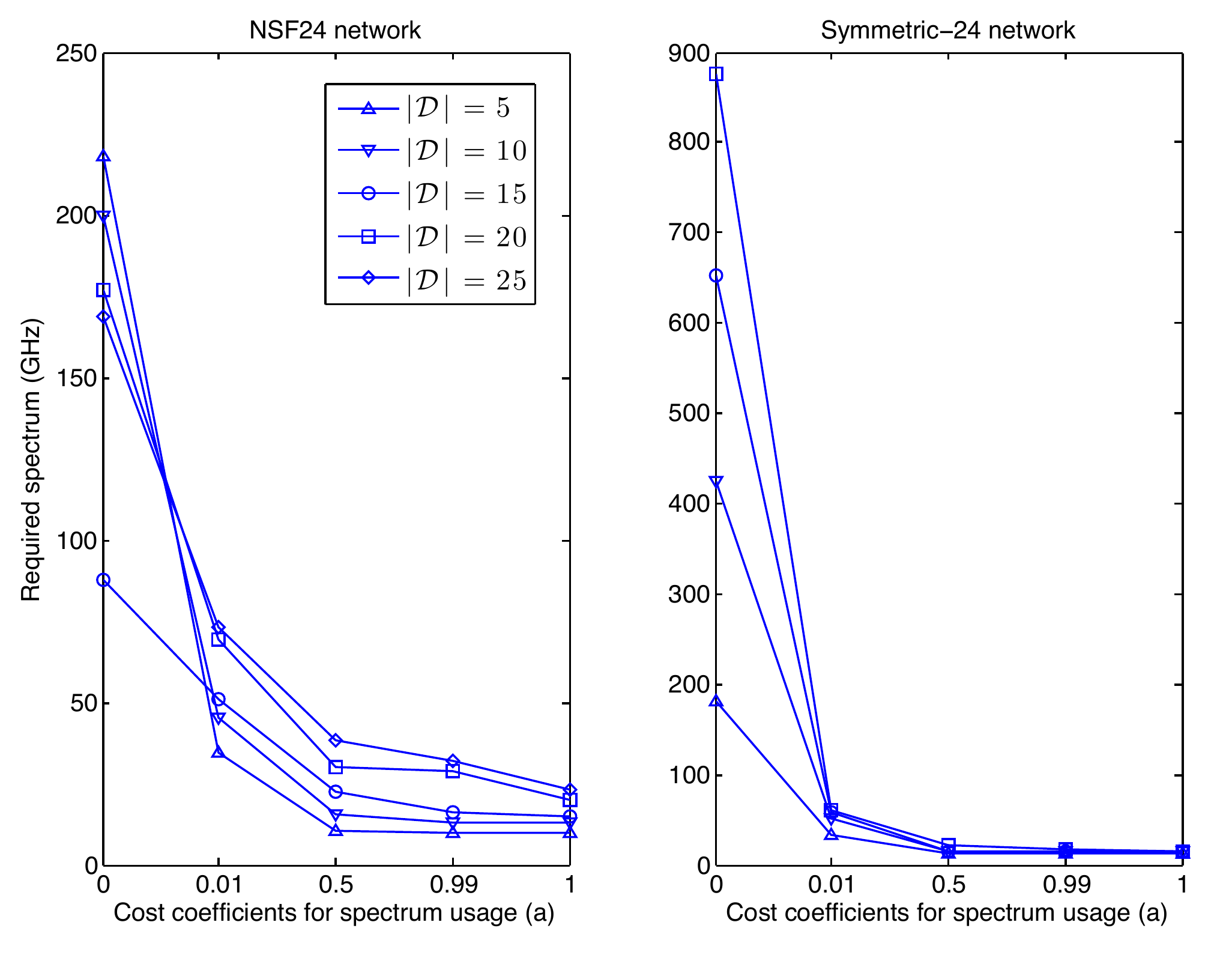}
    \caption{Required spectrum as the coefficient $a$ in the objective function (\ref{eq:multiple}) varies. Note that the horizontal axis is not drawn to scale.}
    \label{fig:coeff_c}
\end{figure}

\begin{figure}[!tb]%[htdp]
   \centering
        \includegraphics[width=3.3in]{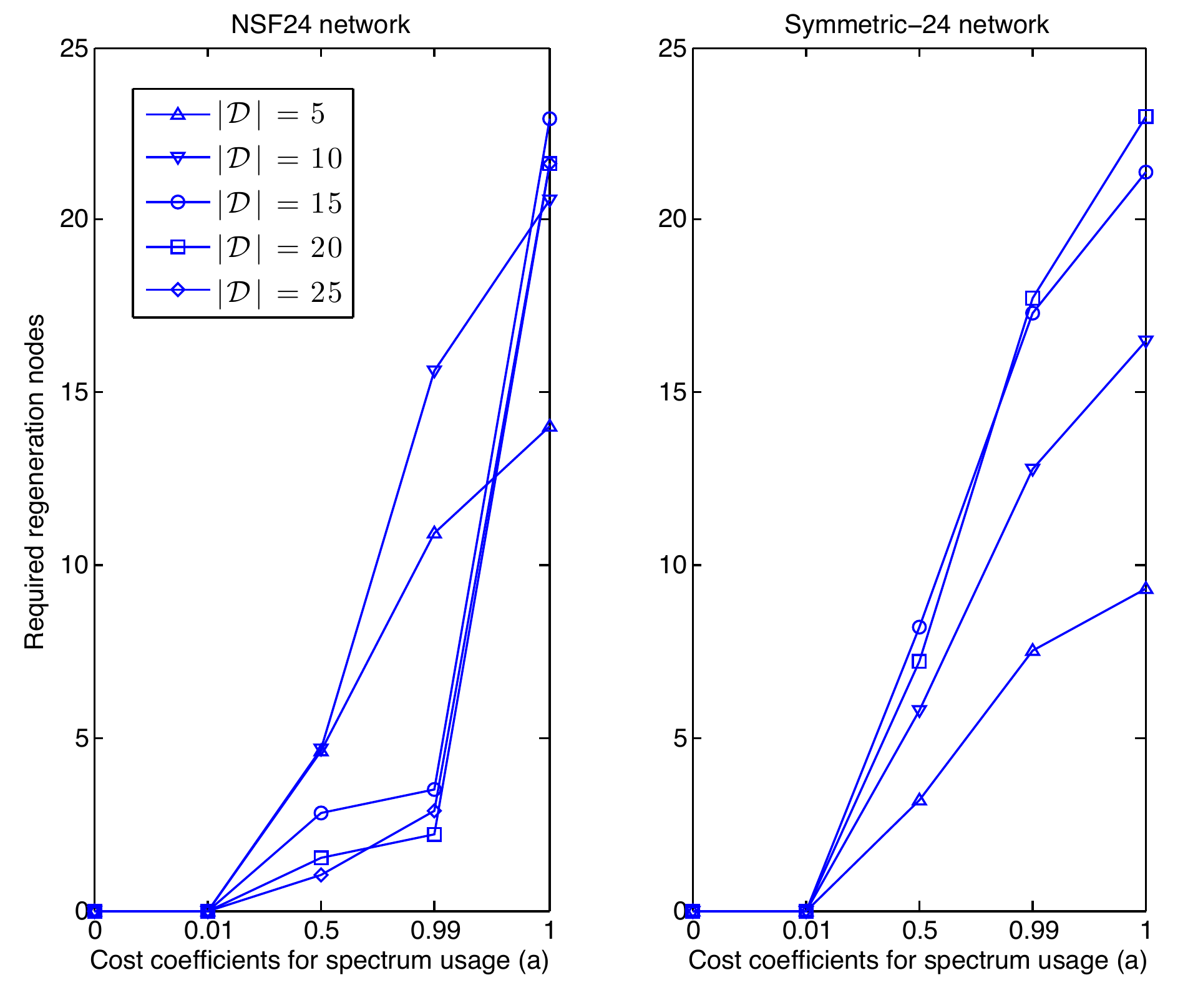}
    \caption{Required number of regeneration nodes as the coefficient $a$ in objective function (\ref{eq:multiple}) varies. Note that the horizontal axis is not drawn to scale.}
    \label{fig:coeff_rn}
\end{figure}

We  investigate the ability of the MILP optimization to effectively trade-off spectral usage with regenerator usage by including the cost of regeneration resources (namely the number of regeneration nodes) in our objective function according to Eq.~(\ref{eq:multiple}).  The resulting spectrum and regeneration node requirement based on different values of the cost coefficient $a$ are shown in Figs.~\ref{fig:coeff_c} and~\ref{fig:coeff_rn}, respectively. We show the effect of the number of demands $|\mathcal D|$. Assuming {\em  a priori} knowledge of the cost relationship between spectrum  and regeneration resources, network designers can choose the cost coefficients accordingly. The results show that when the spectrum cost is not considered (i.e., $a = 0$), the number of regeneration nodes is minimized and the required spectrum is large.\footnote{By configuring the demand volume and spectral efficiency, we make sure there is a spectral efficiency available so that it is possible that no demand requires regeneration. Therefore, in the case of $a = 0$, the number of regeneration nodes is always minimized to zero.} The required spectrum  in this case is also highly irregular, as it is entirely unconstrained. On the contrary, when the regeneration cost is not considered (i.e.: $a = 1$) the required spectrum is minimized but the number of regeneration nodes is high. It is interesting to note that by assigning even a relative small coefficient to regeneration cost (i.e.: $a = 0.99$), we are able to maintain a similar required spectrum but greatly reduce the number of regeneration nodes needed.

%%.. the crossing of results from coeff = 0.01 and 0.99 in network NSF24 is not intuitive..


\subsection{Wavelength and Modulation Scheme Conversion}
\begin{figure}[!tb]%[htdp]
   \centering
        \includegraphics[width=3.3in]{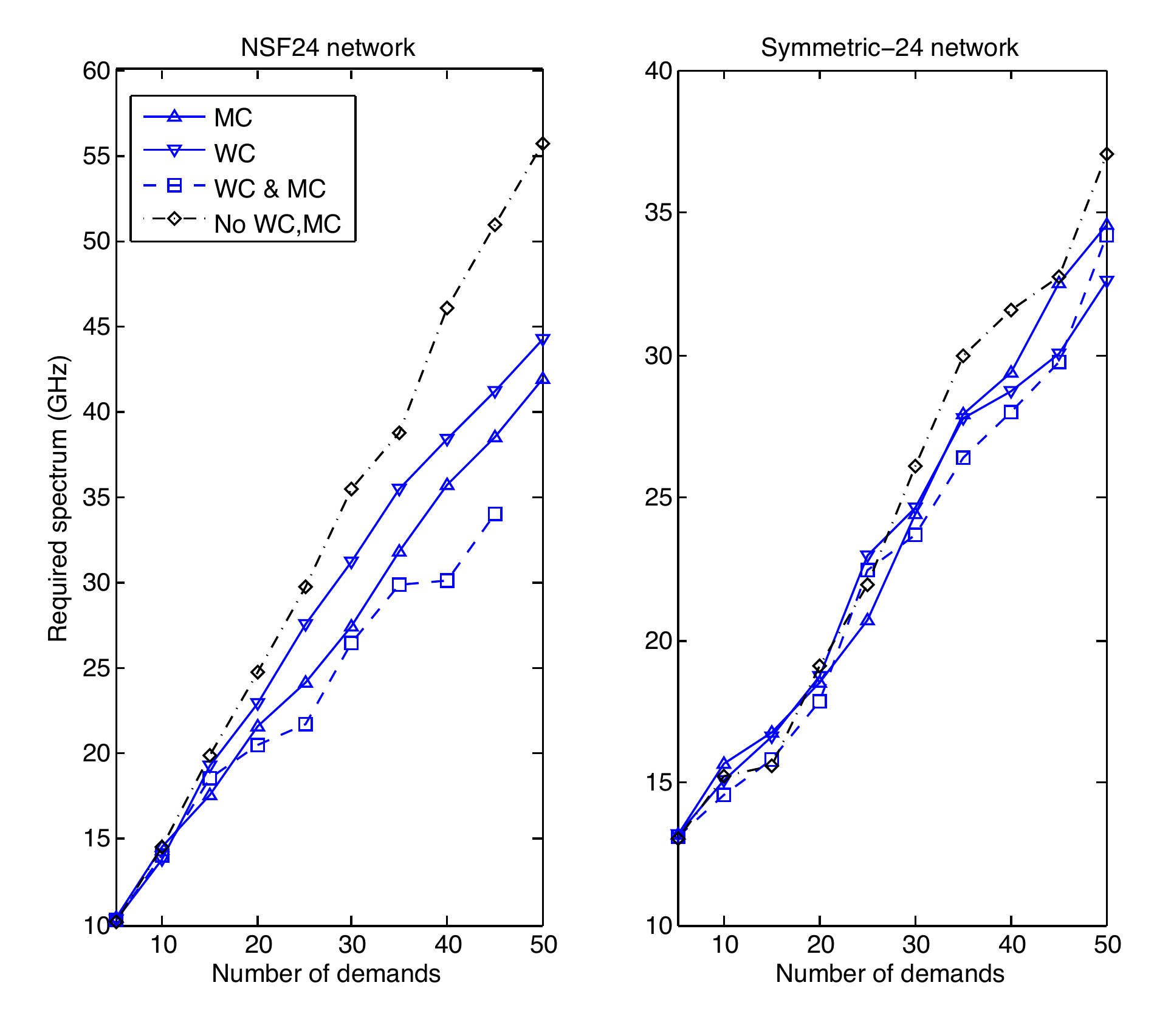}
    \caption{Spectrum usage comparison using the recursive MILP with and without wavelength and/or modulation conversion}
    \label{fig:wavse}
\end{figure}

Since regeneration involves OEO conversion, it can improve the signal quality, so as to extend the TR, and can also provide an opportunity to change the spectrum and modulation assigned starting from that node. We run simulations to show the impact on the spectrum requirements of using the capability to convert the wavelength and/or modulation at the regeneration nodes. In Fig.~\ref{fig:wavse}, we solve the resource allocation problem with our recursive MILP formulation, since the added flexibility of WC and MC increases the complexity of the problem considerably. Both WC and MC reduce the amount of spectrum required to support the traffic demand. WC allows signals to be re-allocated on some links so as to fill gaps left by other traffic demands. The improvement made by wavelength conversion depends on the fragmentation condition of the network. When one link is heavily congested, it then becomes a bottleneck of the network creating unnecessary fragmentation on other links. MC takes advantage of the difference in the length of transparent segments for each lightpath. If they are significantly different, the spectrum saved by optimizing the spectral usage based on each segment becomes significant.  The NSF-24 network has well-documented bottleneck paths, and thus benefits more from WC than a more symmetric topology.  The symmetric network also has equal link lengths, and can therefore not exploit MC as much as the more heterogeneous NSF-24.

In Fig.~\ref{fig:wavse}, when we compare the WC case with the case when both wavelength and modulation scheme conversion are available, we expect the latter to always outperform the former, since it has more flexibility. However, the results (which are averaged over 20 trials) for the symmetric network show that this relationship is not guaranteed. Solving the MILP recursively optimizes the solution in each iteration according to its constraints, yet does not necessarily leads to the optimal solution for the whole traffic matrix when looking at multiple iterations together. Each iteration can only find the local optimum for its sub-problem and the local optimal solution of a sub-problem may not be one part of the global optimal for the whole problem. To put it in terms of resource assignment, in the previous iterations resources may be assigned to demands according to the sub-problem, but such assignment may be not optimal when one considers the whole traffic matrix, which leads to an increased spectrum requirement. The same limitation exists in dynamic real-time RSA where no future traffic information is available. The RSA algorithm cannot assign physical resources to current traffic demands taking unknown future demands into consideration. One could introduce a conservative rule based on long-term statistics that prevents over-assigning physical resources to current demands.

%\begin{figure}[htb]%[htdp]
%\vspace{-.1in}
%   \centering
%        \includegraphics[width=\columnwidth]{p8wcse_c}
%    \caption{Spectrum usage for demands }
%\vspace{-.1in}
%    \label{fig:WAVSE}
%\end{figure}

%\begin{figure}[htb]%[htdp]
%\vspace{-.1in}
%   \centering
%        \includegraphics[width=\columnwidth]{p13limitedRN1}
%    \caption{Spectrum usage with limited number of regeneration nodes}
%\vspace{-.1in}
%    \label{fig:limRN}
%\end{figure}
\subsection{Regeneration Node Placement}
\begin{figure}[!tb]%[htdp]
   \centering
        \includegraphics[width=3.3in]{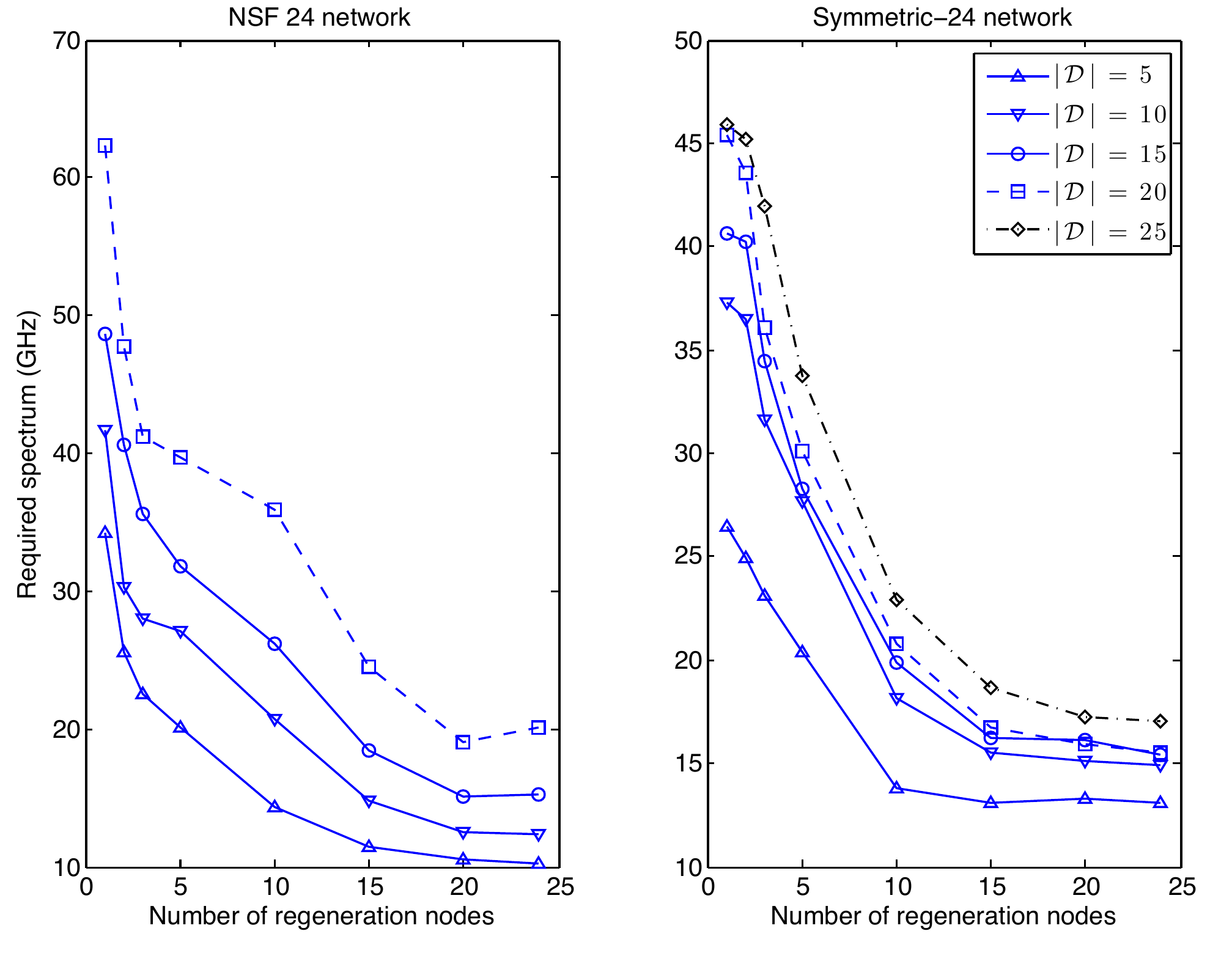}
    \caption{Spectrum usage with a limited number of regeneration nodes}
    \label{fig:limRN}
\end{figure}

When regeneration resources are scarce, careful network planing is important to minimize capital expenditure. In order to show the tradeoff between the number of regeneration nodes and the required spectrum, we simulate a case where the network has a limited number of regeneration nodes. In order to show the change of performance by adding additional regeneration nodes we keep the existing regeneration nodes unchanged once assigned. We use our results in Fig.~\ref{fig:coeff_c} for cost coefficient $a=0.5$ to find and rank the most often used regeneration node locations on average (over 20 trials). After allocating a limited number of nodes as regeneration nodes according to this ranking, we minimize the required spectrum. Our results in Fig.~\ref{fig:limRN} show that for the NSF-24 network, the required spectrum reaches its minimum when 20 nodes have been allocated as regeneration nodes; for the symmetric 24 network, the required spectrum is close to minimum already when the number of regeneration nodes reaches 15. We believe this is due to the symmetric structure of the symmetric 24 node network, where most node pairs share joint intermediate nodes.

%\begin{figure}[htb]%[htdp]
%\vspace{-.1in}
%   \centering
%        \includegraphics[width=\columnwidth]{wavse_recursive30trials60D}
%    \caption{Spectrum usage for demands }
%\vspace{-.1in}
%    \label{fig:WAVSE2}
%\end{figure}

%\begin{figure}[htb]%[htdp]
%\vspace{-.1in}
%   \centering
%        \includegraphics[width=\columnwidth]{recursiveruntime}
%    \caption{run time for solver}
%\vspace{-.1in}
%    \label{fig:runtime}
%\end{figure}

%\begin{figure}[htb]%[htdp]
%\vspace{-.1in}
%   \centering
%        \includegraphics[width=\columnwidth]{wavse_recursive5trials50D24N}
%    \caption{5 trials}
%\vspace{-.1in}
%    \label{fig:recursive}
%\end{figure}
%\begin{figure}[htb]%[htdp]
%\vspace{-.1in}
%   \centering
%        \includegraphics[width=\columnwidth]{wavse_recursive9trials50D24N}
%    \caption{9 trials}
%\vspace{-.1in}
%    \label{fig:recursive}
%\end{figure}
%\begin{figure}[htb]%[htdp]
%\vspace{-.1in}
%   \centering
%        \includegraphics[width=\columnwidth]{wavse_recursive30trials50D24N}
%    \caption{30 trials}
%\vspace{-.1in}
%    \label{fig:recursive}
%\end{figure}

\section{Conclusion}\label{sec:conclusion}
In this paper we propose an MILP formulation to investigate the impact of technologies such as allowing multiple modulation schemes, signal regeneration, wavelength conversion, and modulation conversion on the required spectrum of the EON.  We show through simulation that equipping systems with signal regenerating nodes that control physical impairments and allow for modulation and/or wavelength conversion reduces the amount of spectrum required. Such improvements depend on the topology of the network. We also show the impact of having a limited number of regeneration nodes and different topology structures. In order to balance the optimality and complexity we propose a recursive MILP formulation that yields a suboptimal yet comparable solution to the MILP by using a lower and more consistent running time. The performance of the recursive model depends on factors such as heuristic demand ordering and iteration size.

\section*{Acknowledgment}

This work was supported in part by NSF grants CNS-0915795 and CNS-0916890.

\bibliographystyle{IEEEtran}
\bibliography{Xu_bib}
%,ACOj}

\end{document}